\documentclass[%
 reprint,
 amsmath,amssymb,
 aps,
prstab,
]{revtex4-2}

\usepackage{graphicx}% Include figure files
\usepackage{dcolumn}% Align table columns on decimal point
\usepackage{bm}% bold math
\usepackage[dvipsnames]{xcolor} %NMM
\usepackage{multirow} %NMM https://tex.stackexchange.com/questions/43656/table-with-multiple-merging
\usepackage{colortbl} %NMM https://tex.stackexchange.com/questions/50349/color-only-a-cell-of-a-table
\newcommand{\formatCaseName}[1]{\texttt{#1}}
\begin{document}

\preprint{APS/123-QED}

\title{Single-shot, transverse self-wakefield reconstruction from screen images}% Force line breaks with \\

\author{N. Majernik$^{1,\dagger}$}
%\altaffiliation[Now at SLAC National Accelerator Laboratory]{}
%\email{Now at SLAC National Accelerator Laboratory -- \\ majernik@slac.stanford.edu}
\email{majernik@slac.stanford.edu}

\author{W. Lynn$^1$}
\author{G. Andonian$^1$}

\author{T. Xu$^{2,\dagger}$}
%\altaffiliation[Now at SLAC National Accelerator Laboratory]{}
%\email{Now at SLAC National Accelerator Laboratory.}

\author{P. Piot$^{2,3}$}

\author{J. B. Rosenzweig$^1$ \\ \phantom{.}}

\affiliation{
$^1$University of California Los Angeles, Los Angeles, California 90095, USA
}

\affiliation{ 
 $^2$Northern Illinois University, DeKalb, Illinois 60115, USA
 }

\affiliation{ 
 $^3$Argonne National Laboratory, Lemont, IL 60439, USA
 }

\affiliation{ 
 $^\dagger$Now at SLAC National Accelerator Laboratory, Menlo Park, California 94025, USA
 }

\date{\today}% It is always \today, today,

\begin{abstract}
A single-shot method to reconstruct the transverse self-wakefields acting on a beam, based only on screen images, is introduced. 
By employing numerical optimization with certain approximations, a relatively high-dimensional parameter space is efficiently explored to determine the multipole components of the transverse-wakefield topology up to desired order. 
The reconstruction technique complements simulations, which are able to directly describe the wakefield composition based on experimental conditions. 
The technique is applied to representative simulation results as a benchmark, and also to experimental data on wakefield observations driven in dielectric-lined structures.

%\begin{description}
%\end{description}
\end{abstract}

\maketitle

%\tableofcontents

\section{INTRODUCTION}

Electromagnetic wakefields generated by relativistic particle beams propagating in slow-wave structures or near resistive boundaries are an important consideration in many aspects of accelerator and beam physics research.
%, may be desirable and intentionally excited 
%Transverse self-wakefields are an important consideration in many facets of beam physics, arising in dielectric, metallic, and plasma structures. 
The transverse components of self-generated wakefields that act back on the beam may be desirable and intentionally excited, \textit{e.g.}, as passive streakers for beam diagnostics \cite{bettoni2016} or in emerging wakefield-based focusing schemes \cite{lynnIPAC,lynnNAPAC}. 
However, transverse self-wakefields may also lead to emittance growth and possibly seed a beam-breakup instability~\cite{li2014, oshea2020}.
Detailed information about the wakefields and the subsequent beam dynamics are thus critical for accelerator design and, in particular, advanced acceleration concepts based on wakefields in plasma or structures.
Due to the spatial and temporal scales involved, as well as the  relativistic nature of the beam and its induced fields, it is challenging to directly interrogate self-wakefields \textit{in situ}.
Nevertheless, it is possible to ascertain properties of the electromagnetic wakefields based on the measured frequency content of the outcoupled radiation \cite{oshea2020,majernik2022}, or by comparing experimental measurements of beam properties to simulations. However, a method to directly determine the detailed characteristics of wakefields themselves is much more desirable. 
In a plasma wakefield accelerator, it may be possible to directly image the plasma wave and thus infer information about the wakefields either optically \cite{marques1996temporal, siders1996laser, matlis2006snapshots, buck2011real, savert2015direct} or using a secondary electron beam \cite{zhang2016capturing}. %Maybe look at https://journals.aps.org/rmp/pdf/10.1103/RevModPhys.90.035002 if more examples are desired
Such non-destructive approaches are not applicable to wake interactions which occur in vacuum structures within material enclosures (\textit{e.g.} wakefields from dielectric lined or corrugated metallic structures).
Other wakefield-based techniques have been proposed which may be more appropriate for in-vacuum interactions, including using tailored probe beams \cite{halavanau2021}. 
In this paper we introduce a new technique for investigating the transverse self-wakefields without requiring any additional experimental infrastructure, by measuring, on a standard screen, the two-dimensional transverse projections (in the $(x,y)$ plane) of the beam undergoing transverse self-wake interactions. %Maybe discuss some of the more direct plasma diagnostics; explain that in a plasma, physical matter corresponds to fields. Explain why this falls apart for in vacuo interactions.

In order to develop a foundation for the reconstruction approach, we revisit the formalism of wakefield descriptions using a standard multipole decomposition.
It is possible to describe the transverse wakefields, $\bm{W}$, expressed in terms of the net transverse momentum kick, at each longitudinal position along the beam with a  potential-based formalism \cite{tanabe2005}:
\begin{equation}
\begin{aligned}
&A + i V = \sum_{n=1}^{\infty} (a_n + i b_n)(x + i y)^n, \\ &\bm{W} = -\left \{\frac{\partial V}{\partial x}, \frac{\partial A}{\partial x} \right \}.
\end{aligned}
\label{eq:multipoleDef1}
\end{equation}

\noindent This yields transverse wakefields of the form

\begin{equation}
\begin{aligned}
    \bm{W} &= -\left \{\frac{\partial V}{\partial x}, \frac{\partial A}{\partial x} \right \}  = \\ &\{ -b_1 - 2 a_2 y - 2 b_2 x -6 a_3 x y  + 3 b_3 (-x^2 + y^2) + ..., \\  &-a_1 - 2 a_2 x + 2 b_2 y + 3 a_3 (-x^2 + y^2) + 6 b_3 x y + ... \},
\end{aligned}
\end{equation}

\noindent where $a_n$ and $b_n$ are the $n^\mathrm{th}$-order skew and normal multipole coefficients respectively. By confining the reconstruction to low order multipole moments (an approximation that is quantified and justified below) it is possible to infer the self-wakefields by examining the change that they cause in the transverse distribution of a beam imaged on a screen.

Furthermore, in this work we utilize a thin-lens approximation, \textit{i.e.} that the beam spatial distribution does not change significantly over the length of the wakefield interaction itself, and thus only the integrated kick from the interaction (in the instantaneous approximation) is reconstructed. 
%\textcolor{red}{Need to find a more careful way to phrase this. Strictly, does not not need to be longitudinally invariant; e.g. a corrugated structure is fine. Just need the length scale of the spatial evolution of the beam to be $>>$ the length scale of those changes. Basically that it's OK for a range of wakefields as long as we can treat the problem as an average of these. Can even apply to, e.g. AGWA, if this condition is satisfied but it may not be illuminating. Maybe refer to the notion of "integrated kick"?}. [to go along this comment it may be good to add the mathematical condition -- we a e basically stating that the integral over the length of the interaction can be replaced by the interaction length times an average kick]

\begin{figure*}[ht] 
\centering
\includegraphics[width=1.0\linewidth]{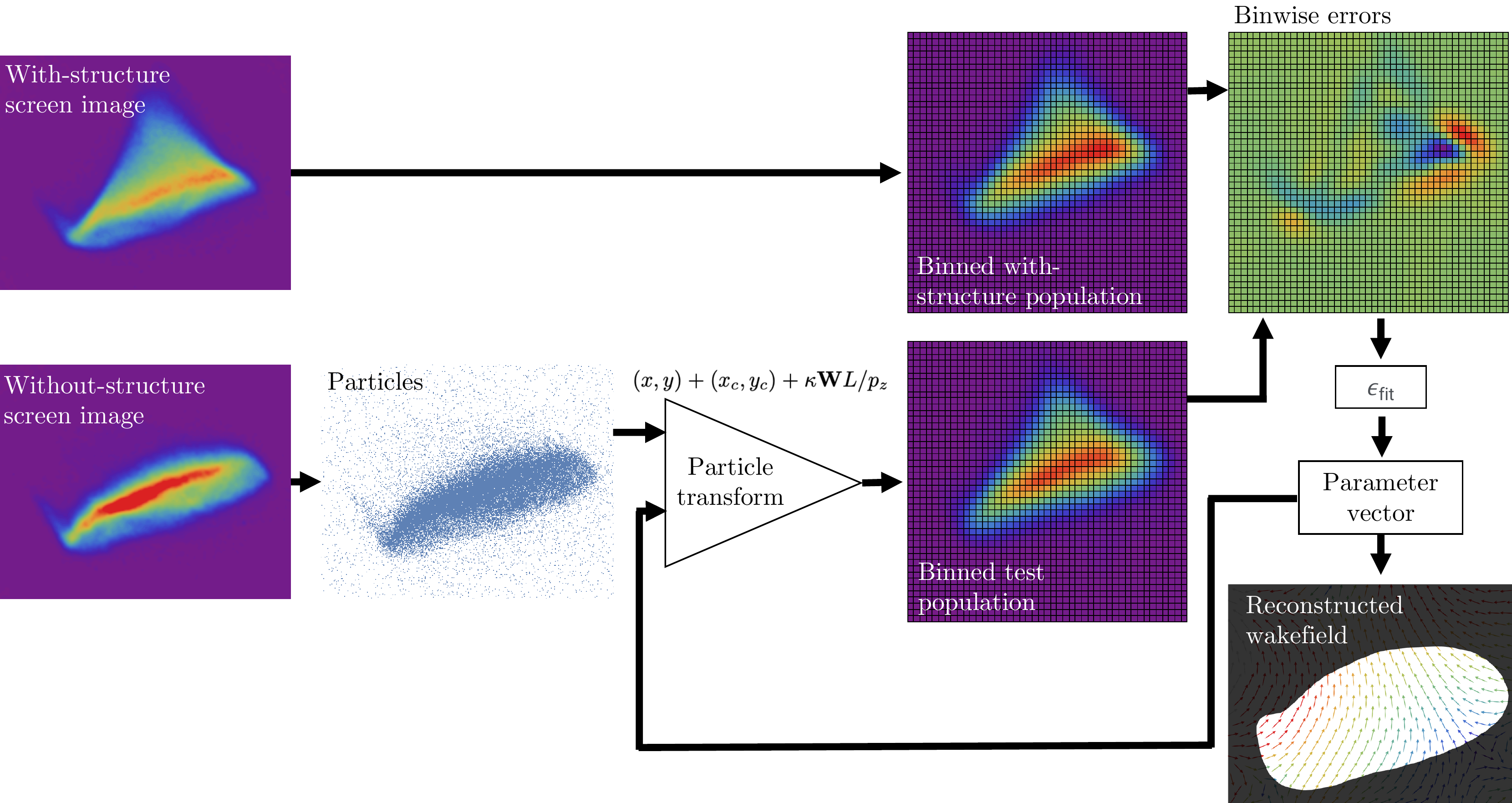}
\caption{Schematic overview of the algorithm for the single-shot reconstruction of transverse beam wakes from screen images. The bin size has been exaggerated for clarity.}
\label{fig:flowChart}
\end{figure*}

We will also assume that the transverse wakefield at every longitudinal position along the beam (denoted simply as $z$, with beamline distances indicated by $s$) is proportional up to a multiplicative constant, $\kappa(z)$, which is a normalizing factor determined from the data. 
The approximation holds provided that the bunch length is short compared to the wavelength of all modes which appreciably contribute to the overall wakefield and will be further justified below. 
In this limit, each mode's convolution with the beam, over the duration of the beam, is well-approximated as a curve which is monotonically  increasing towards the tail of the beam. 
Though each of these contributions will have different slopes depending on their wave-number, since they all start at zero (by construction) their ratios will be constant, justifying this approximation. 

In this paper, we first describe the mathematical algorithm to reconstruct transverse self-wakefields. 
We then use the algorithm to reconstruct wakefields using simulation results as a benchmark. 
The technique is then extended to experimental data from a recent dielectric wakefield study conducted at the Argonne Wakefield Accelerator (AWA) facility. 
Finally, we discuss the range of applicability for the reconstruction and some simple ways to extend the technique's  fidelity.

\section{Transverse wakefield reconstruction}

The goal of the reconstruction method (Fig.~\ref{fig:flowChart}) is to fully describe the transverse self-wakefields affecting the beam.
The input for the technique is two images, taken on the same screen downstream of the wakefield interaction point: one with the structure in place, and one without. Herein, ``structure'' is used generically and can refer to either a plasma, dielectric-lined waveguide, corrugated waveguide, or any other wakefield-generating element that acts back on the beam. 
The screen images are used as probability density functions (PDF) which are sampled to create two populations of $n$, $(x,y)$ macroparticles. 
A 2D grid of bins is defined which is large enough to cover both beam distributions, with a bin spacing small enough to resolve the beams' structures. 
The wakefield-affected (\textit{i.e.} ``with-structure'') macroparticles are binned, yielding bin-wise counts $B_{i,j}$. 
The without-structure macro-particles are then transformed according to 
%$(x,y) + (x_c, y_c) + \frac{\kappa \mathbf{W} L}{p_z}$ 
$(x,y) + (x_c, y_c) + \kappa \mathbf{W} L/p_z$ 
where $(x_c, y_c)$ are corrections for beam drift between the without-structure and with-structure shots (omitted for simulations where this effect is not present), $\kappa$ is the wakefield amplitude scale factor, sampled independently from the PDF of the random variable $\mathcal{K}$, $L$ is the effective drift length from the structure to the screen, and $p_z$ is the mean longitudinal momentum of the beam. 
The transformed macroparticles are binned, yielding $\widetilde{B}_{i,j}$. We define the overall error as the sum of the absolute errors for all the bins, divided by twice the number of macroparticles to avoid double-counting: %$\sum{|\widetilde{B}_{i,j}-B_{i,j}|/n}$. 

\begin{equation}
\epsilon_\mathrm{fit} = \sum_{i,j}{\frac{|\widetilde{B}_{i,j}-B_{i,j}|}{2 n}} ,
\label{eq:optimizerError}
\end{equation}

\noindent which ranges from zero to one. 
A numerical optimizer is employed to minimize this error term by finding the best parameter vector $\{x_c, y_c, \mathcal{K}_1, \mathcal{K}_2, ..., a_1, b_1, a_2, b_2, ...\}$ where $\mathcal{K}_1, \mathcal{K}_2, ...$ parameterize $\mathcal{K}$, with multipole coefficients $a_i, b_i$ up to the selected order. 
The technique is visually summarized in Fig.~\ref{fig:flowChart}.

%\begin{figure}[H] 
%\centering
%\includegraphics[width=0.88\linewidth]{Figures/flowChart}
%\caption{\textcolor{red}{PLACEHOLDER} Schematic overview of the algorithm for the single-shot reconstruction of transverse beam wakes from screen images.}
%\label{fig:flowChart}
%\end{figure}

The random variable $\mathcal{K}$ results from weighting the probability of the wakefield strength at a longitudinal position, $\kappa(z)$, by the current at that position, $I(z)$. 
Depending on the available information and the expected complexity, different definitions for the PDF may be used. 
For the simulated cases below, the values of $\kappa(z)$ and $I(z)$ can be found directly and used to construct the PDF. 
For experiments for which there are trusted simulations, this approach may also be used. 
However, it is also possible to use less prior knowledge. Shown is an example of a convenient and versatile PDF, used by several reconstructions below, with one optimizer parameter, $\mathcal{K}_1$: %$f_\mathcal{K}(\kappa) = \begin{cases} (1+ \mathcal{K}_1) \kappa^{\mathcal{K}_1} & 0 < \kappa < 1 \\ 0 & \mathrm{otherwise} \end{cases}$.
\begin{equation}
    f_\mathcal{K}(\kappa) = \begin{cases}
   (1+ \mathcal{K}_1) \kappa^{\mathcal{K}_1} & 0 < \kappa < 1 \\
   0 & \mathrm{otherwise}
\end{cases}.
\label{eq:1DPDF}
\end{equation}
If this does not give convergent results, it is possible to define the distribution based on other assumptions or with a greater number of free parameters.

Several methods were attempted for optimization in this relatively high dimensional parameter space. Derivative-based optimizers encountered difficulties due to the discontinuous and noisy nature of the data, resulting from binning discrete particles. 
Bayesian optimization methods performed better, but the comparatively high computational overhead of this technique made it suboptimal. 
The best performance was found using a derivative-free \cite{koziel2011} global optimization, in particular, differential evolution. This approach is not guaranteed to arrive at the complete wakefields, but since the relevant wakefields are often dominated by a few, low degree multipole components, it serves as an efficient and effective approximation.
Alternative options may include, instead of binning the particles to discrete bins, using a differentiable optimization based on kernel density estimation \cite{roussel2022phase}.

\section{Benchmarking to simulation}

A variety of dielectric wakefield scenarios were simulated using the finite-difference time-domain (FDTD) particle-in-cell (PIC) code WARP \cite{WARP} (summarized in the Appendix, in Table \ref{tab:resultsSummary}). These scenarios are inspired by recent experimental work which investigated wakefields arising from a variety of beam distributions, including those containing skew components. 

\begin{figure}[!] 
\centering
\includegraphics[width=1.0\linewidth]{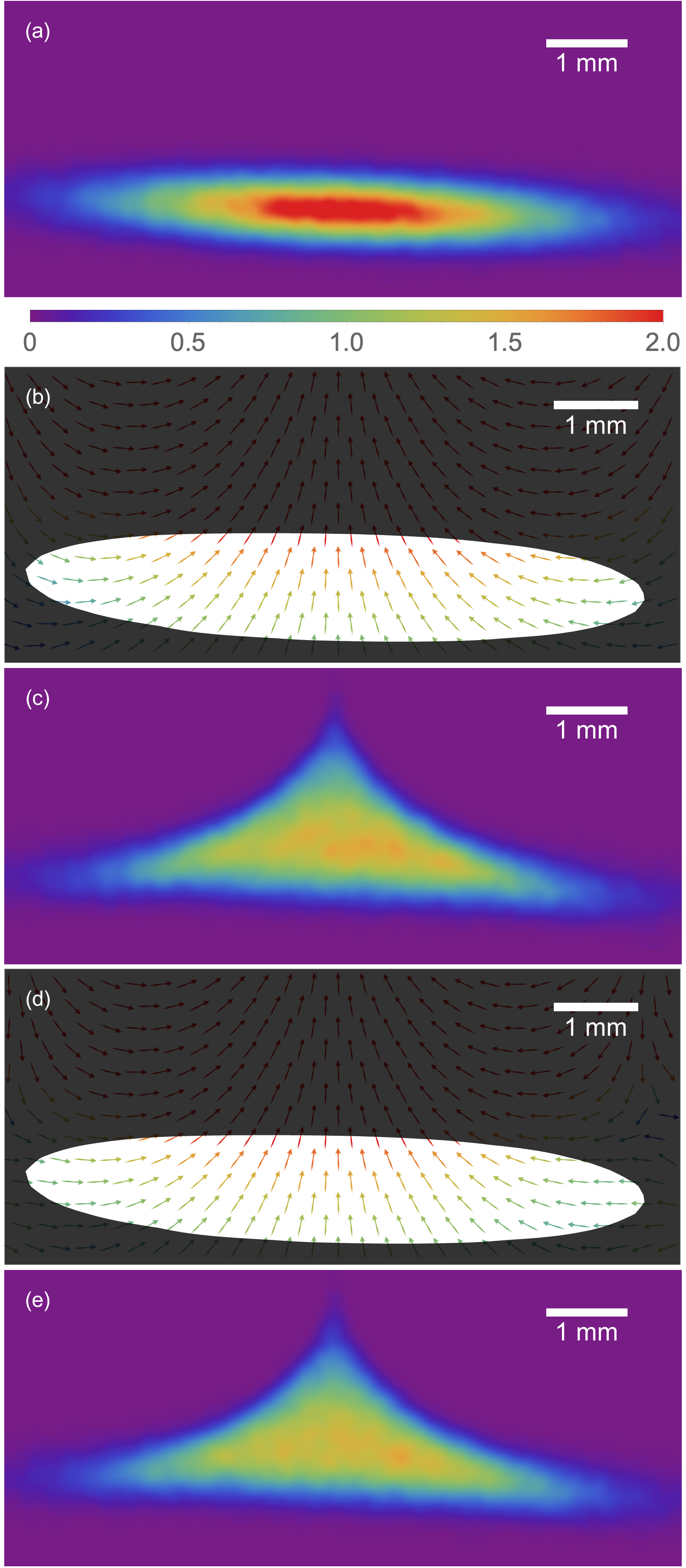} 
\caption{ Figures relating to the \formatCaseName{500~MeV, top slab, with PDF} WARP simulation and reconstruction (see Table \ref{tab:resultsSummary}) \textbf{(a)} Transverse distribution of without-structure beam. (Plot range is constant for all subfigures. Charge density [Arb. units] color scale bar is constant between distributions). \textbf{(b)} Reconstructed wakefields, $\bm{W}_\mathrm{opt}$. White region contains 90\% of beam charge. \textbf{(c)} Implied with-structure beam distribution by $\bm{W}_\mathrm{opt}$. \textbf{(d)} Ground truth wakefields, $\bm{W}_\mathrm{agg}$. \textbf{(e)} Ground truth with-structure beam distribution.}
\label{fig:simulationPrimaryReconstructionSequence}
\end{figure}

In the first scenario chosen, a beam with a charge of 1.95 nC, bunch size $\sigma_x$ = 1.76~mm, $\sigma_y$ = 0.23~mm, $\sigma_z$ = 0.64~mm, and a tilt of 2.2$^{\circ}$ (see Fig.~\ref{fig:simulationPrimaryReconstructionSequence}) was propagated past an alumina slab ($\epsilon_r =$ 9.9) 5 mm thick and 150~mm long, coated on the side opposite the beam with a layer of metal. 
The beam centroid was 0.91~mm below the surface of the dielectric. This simulation is most similar to the recent wakefield experiment performed at AWA \cite{lynnSkew} with the exception that the simulated beam energy was 500~MeV. This permitted us to freeze the motion of the beam during the interaction to ensure the validity of the thin lens assumption. 
The transverse momentum of each particle was recorded before and after the simulated interaction and used to construct an interpolated function for the transverse wakefield $\bm{W}_\mathrm{sim}(x, y, z)$, taken to be the ground truth in the analysis.

$\bm{W}_\mathrm{sim}(x, y, z)$ is then sampled at a number of $z$ slices, spaced with $\Delta z \ll \sigma_z$. These fields were fit with the best $\bm{W}_\mathrm{guess}$ by finding the vector $\bm{c}_\mathrm{slice} = \{a_1, b_1, a_2, b_2, ..., a_n, b_n\}$ which minimizes
\begin{equation}
    %{\mathrm{mean}_{95\%} } {\left ( \left | \bm{W}_\mathrm{sim}(x_k, y_k, z_\mathrm{slice})-\bm{W}_\mathrm{guess}(x_k, y_k) \right | \right )},
    \left \langle \left | \bm{W}_\mathrm{sim}(x_k, y_k, z_\mathrm{slice})-\bm{W}_\mathrm{guess}(x_k, y_k) \right | \right \rangle_{95\%} ,
\label{eq:simError}
\end{equation}
\noindent where $k$ counts along the simulation macro-particles and $\left \langle \cdot \right \rangle_{95\%} $ is a truncated mean which excludes 5\% of the population with the worst errors to prevent stray particles from unduly affecting the result. %And takes advantage of the fact that all the transverse slices of the beam are the same up to current
In this case, coefficients up to $n=6$ (dodecapole) were used. %The median of all the relative field errors is used as the figure of merit to avoid the outsized impact of points far from the center.
%All of the errors from these fits were weighted according to the currents, $I(z_\mathrm{slice})$, and the mean relative error of the multipole fits was 12.6\%. Fitting to higher degrees will yield lower errors.
Using these fit coefficients for each slice and then linearly interpolating between the slices yields $\bm{W}_\mathrm{slicewise}$. 
As a confirmation that a high enough multipole order was appropriate for the fit, the slicewise coefficients are compared back to the ground truth wakefield:
\begin{equation}
\begin{aligned}
   &\epsilon_\mathrm{slicewise} = \\ & \frac{\left \langle \left | \bm{W}_\mathrm{sim}(x_k, y_k, z_k)-\bm{W}_\mathrm{slicewise}(x_k, y_k, z_k) \right | \right \rangle_{95\%}}{\left \langle \left | \bm{W}_\mathrm{sim}(x_k, y_k, z_k)\right | \right \rangle_{95\%}} .
\end{aligned}
\label{eq:slicewiseError}
\end{equation}
\noindent Eq.~\ref{eq:slicewiseError} yields a relative error of 1.5\% indicating that, for this case, $n = 6$ is sufficient to describe the transverse wakefields. $\epsilon_\mathrm{slicewise}$ will improve monotonically with greater $n$ but with diminishing returns.

In order to find the \emph{aggregate multipole} coefficients, $\bm{c}_\mathrm{agg}$, calculate the average of all the $\bm{c}_\mathrm{slice}$ vectors, weighted by the current at that longitudinal position, $I(z_\mathrm{slice})$
\begin{equation}
    \bm{c}_\mathrm{agg} = \sum_\mathrm{slices} I(z_\mathrm{slice}) \bm{c}_\mathrm{slice} \bigg/ \sum_\mathrm{slices} I(z_\mathrm{slice}).
    \label{eq:aggmultipoles}
\end{equation}
These coefficients, via Eq.~(\ref{eq:multipoleDef1}), yield the aggregate wakefield $\bm{W}_\mathrm{agg}$. 

\begin{figure}[h] 
\centering
\includegraphics[width=1\linewidth]{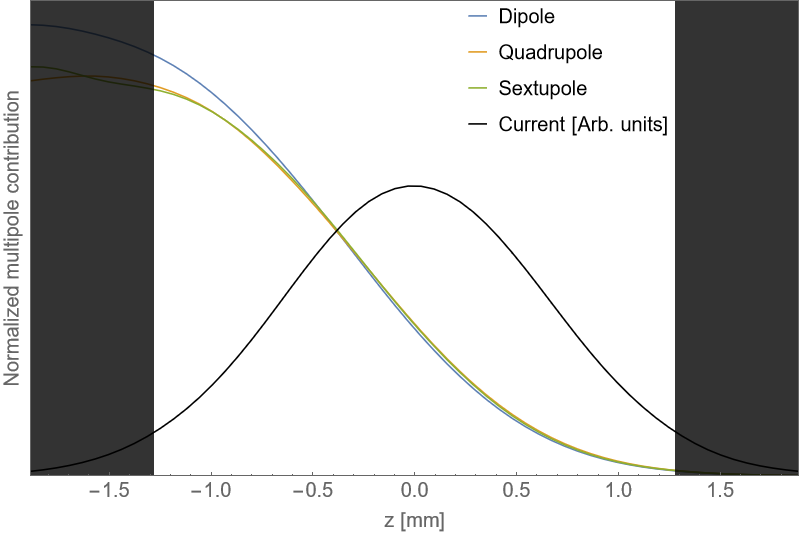}
\caption{At each longitudinal position along the beam, a multipole decomposition is performed. The values of $\sqrt{a_i^2 + b_i^2}$ are shown, after being normalized. Also shown is the current profile of the beam, with the white region near the center corresponding to 90\% of the beam charge. }
\label{fig:slicewiseKicks}
\end{figure}

As previously discussed, we assume that the fields at every slice are equal to the aggregate wakefield, including a multiplicative constant: $\bm{c}_\mathrm{slice} \approx \kappa(z)  \bm{c}_\mathrm{agg}$. 
The approximation is valid if the wavelengths of modes which produce relevant multipoles are long relative to $\sigma_z$, which is reasonable for the scenarios under consideration. 
In Fig.~\ref{fig:slicewiseKicks}, the normalized, slice-wise coefficients for the first three multipole moments are shown as a function of $z$. Since the lines for the dipole, quadrupole, and sextupole orders are approximately the same shape within the area of most of the beam charge, this is evidence that approximating the fields as constant, up to a $z$-dependent multiplicative factor, is reasonable. $\kappa(z)$ is calculated by finding the value which minimizes the difference between $\kappa(z_\mathrm{slicewise}) \bm{W}_\mathrm{agg}$ and $\bm{W}_\mathrm{slicewise}$ at each $z_\mathrm{slice}$ and linearly interpolating between these values. 

\begin{figure}[tb] 
\centering
\includegraphics[width=1\linewidth]{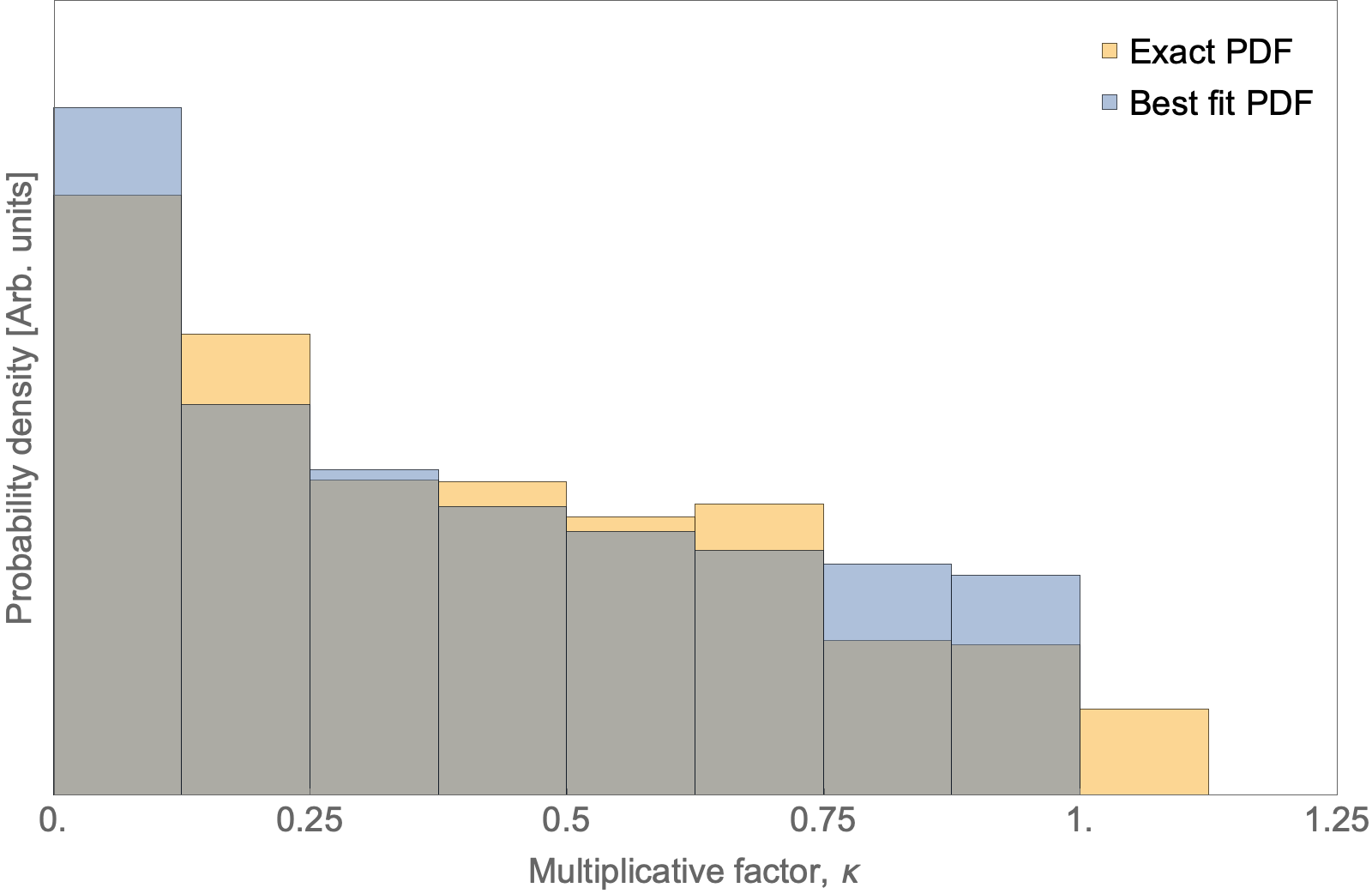}
\caption{A histogram showing the probability density function of $\kappa$, the multiplicative weighting factor. These results are for the \formatCaseName{500 MeV, top slab} WARP simulation with the ``Exact PDF" corresponding to using the $I(z)$ weighted exact $\kappa(z)$ and the ``Best fit PDF" corresponding to Eq.~\ref{eq:1DPDF} with $\mathcal{K}_1 = 1.54$, determined using the reconstruction algorithm.}
\label{fig:kickFactorPDF}
\end{figure}

As a final check of both sufficient multipole order and the $\bm{c}_\mathrm{slice} \approx \kappa(z)  \bm{c}_\mathrm{agg}$ assumption,  $\kappa(z)  \bm{c}_\mathrm{agg}$ is compared to the ground truth at all the macroparticle positions:
\begin{equation}
\begin{aligned}
   &\epsilon_\mathrm{agg} =  & \frac{\left \langle \left | \bm{W}_\mathrm{sim}(x_k, y_k, z_k)-\kappa(z_k)\bm{W}_\mathrm{agg}(x_k, y_k) \right | \right \rangle_{95\%}}{\left \langle \left | \bm{W}_\mathrm{sim}(x_k, y_k, z_k)\right | \right \rangle_{95\%}} .
\end{aligned}
\label{eq:aggError}
\end{equation}
Eq.~\ref{eq:aggError} yields an overall relative error of 2.3\%. Note that $\epsilon_\mathrm{slicewise}$ (1.5\% in this case) will set a lower bound on $\epsilon_\mathrm{agg}$; the close agreement between these values indicate that the longitudinal approximation is reasonable. Specifically, we have managed to capture the overwhelming majority of the information regarding the 3D map of the transverse wakes ($\mathbf{W} : \mathbb{R}^3 \rightarrow \mathbb{R}^2$) using only 12 coefficients for the multipoles and a 1D function, $\kappa(z)$. The PDF for the wakefield strength was constructed from $\kappa(z)$ and $I(z)$; it is shown in Fig.~\ref{fig:kickFactorPDF}.

Simulated screen images, 3 meters downstream of the exit of the structure, were produced from the simulation data and fed into the optimizer described in the previous section. Initially, the kick PDF was provided so the only free parameters were the multipole coefficients which were permitted up to $n=6$. Fig.~\ref{fig:simulationPrimaryReconstructionSequence} shows the without-structure beam and reconstructed with-structure beam and associated fields, as well as the ground truth with-structure beam and fields. The optimizer figure of merit, $\epsilon_\mathrm{fit}$, converged to 1.3\%. Finally, the ground truth aggregate wakefield, $\bm{W}_\mathrm{agg}$, is compared to the reconstructed wakefield, $\bm{W}_\mathrm{opt}$:
\begin{equation}
\begin{aligned}
%  &\epsilon_\mathrm{recons} = & \frac{\left \langle \left | \bm{W}_\mathrm{agg}(x_k, y_k)-\bm{W}_\mathrm{opt}(x_k, y_k) \right | \right \rangle_\mathrm{median}}{\left \langle \left | \bm{W}_\mathrm{agg}(x_k, y_k)\right | \right \rangle_{95\%}},
&\epsilon_\mathrm{recons} = & \left \langle \frac{\left | \bm{W}_\mathrm{agg}(x_k, y_k)-\bm{W}_\mathrm{opt}(x_k, y_k) \right | }{\left \langle \left | \bm{W}_\mathrm{agg}(x_k, y_k)\right | \right \rangle_{95\%}} \right \rangle_\mathrm{median},
\end{aligned}
\label{eq:reconstructionError}
\end{equation}
where $\left \langle \cdot \right \rangle_\mathrm{median} $ is the median. For this scenario, the reconstruction technique described here recovered the ground truth wakefield to within 2.7\%.

\begin{figure}[ht!] 
\centering
\includegraphics[width=0.82\linewidth]{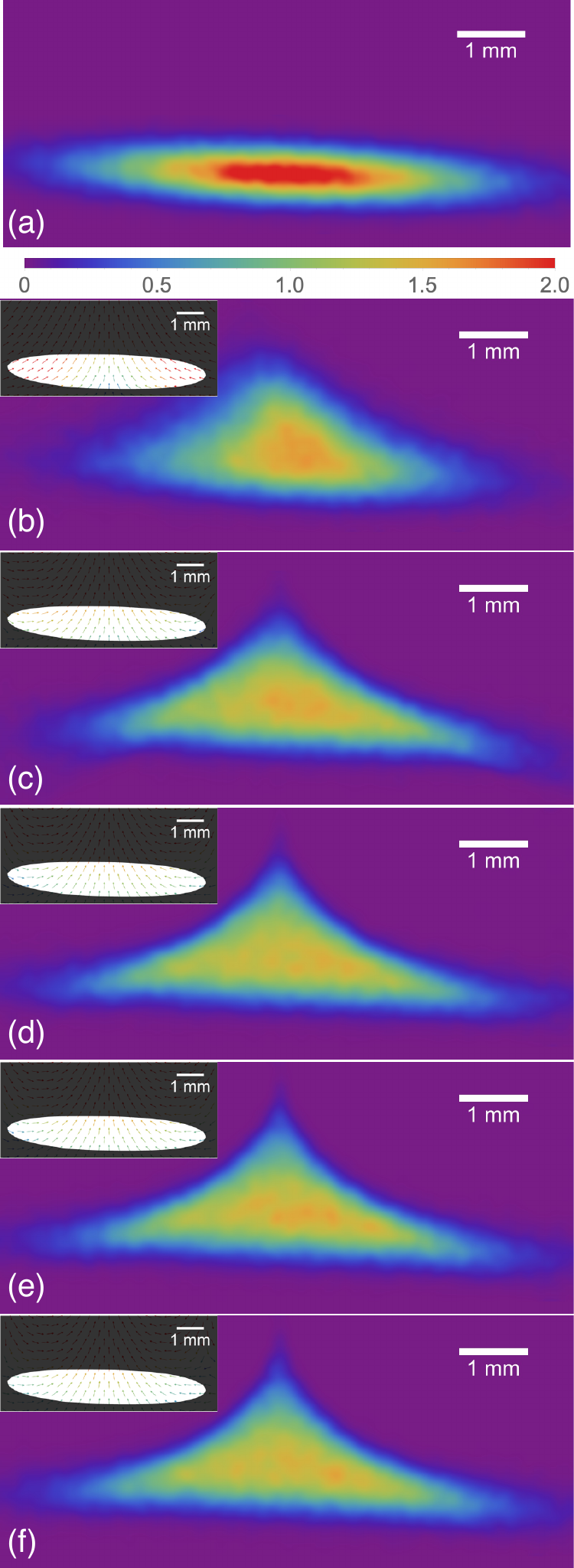}
\caption{Reconstruction of \formatCaseName{500 MeV, top slab} WARP simulation under different reconstruction conditions. Insets show the wakefields that result in the corresponding beam. For additional details refer to Table~\ref{tab:resultsSummary}. \textbf{(a)} Without-structure beam. \textbf{(b)} Not given $\kappa$ PDF, reconstruction to $n=2$. \textbf{(c)} Not given $\kappa$ PDF, reconstruction to $n=4$. \textbf{(d)} Not given $\kappa$ PDF, reconstruction to $n=6$. \textbf{(e)} Given $\kappa$ PDF, reconstruction to $n=6$. \textbf{(f)} Ground truth, with-structure beam. }
\label{fig:simulationReconstructionConditions}
\end{figure}

This particular reconstruction assumes that the PDF of $\mathcal{K}$ is known. If there are accurate simulations available for the experiment under consideration, this is of course possible. More generally though, we anticipate that the $\mathcal{K}$ will not be known. We repeat the reconstruction of this scenario with one more free parameter, $\mathcal{K}_1$, and the PDF from Eq.~\ref{eq:1DPDF}. The final result is quite similar; the reconstructed PDF is shown compared to the ground truth PDF in Fig.~\ref{fig:kickFactorPDF} and the with-structure beam is shown in Fig.~\ref{fig:simulationReconstructionConditions}. Its overall reconstruction error, $\epsilon_\mathrm{recons}$, is degraded only slightly, to 3.7\%.

\begin{figure}[tb] 
\centering
\includegraphics[width=\linewidth]{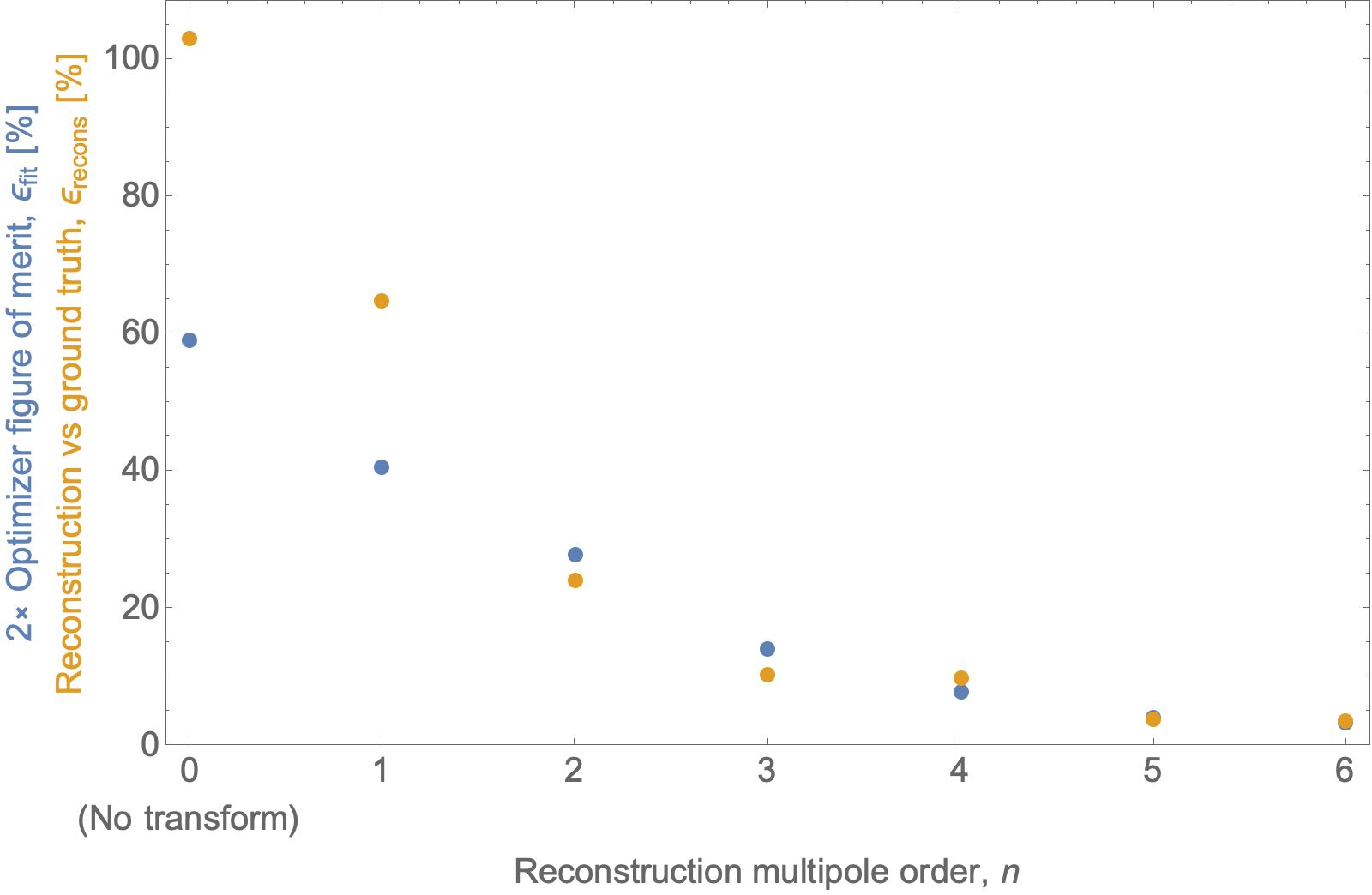}
\caption{The effect of changing the reconstruction order, $n$, on the error metrics $\epsilon_\mathrm{fit}$ and $\epsilon_\mathrm{recons}$ for the \formatCaseName{500 MeV, top slab} WARP simulation. Diminishing returns for both metrics are achieved around $n$ = 4 or 5.}
\label{fig:simulationReconstructionQuality}
\end{figure}

Since the ground truth to which the reconstructions are compared has been defined to finite order, by construction, there is no advantage to giving the reconstruction algorithm additional, free multipole coefficients beyond $n = 6$. However, for experimental data, it will not be known \emph{a priori} to what order reconstruction should be performed. Instead, the number of free parameters should be chosen such that there are diminishing returns for a higher order. This process has been tested on simulated data and summarized in Fig.~\ref{fig:simulationReconstructionQuality}. Here it can be observed that both optimizer figure of merit $\epsilon_\mathrm{fit}$ and the overall reconstruction error $\epsilon_\mathrm{recons}$ trend down with increasing reconstruction order but reach diminishing returns around $n$ = 4 or 5. The effect of changing the conditions of the reconstruction, including varying $n$ can be visually inspected in Fig.~\ref{fig:simulationReconstructionConditions}. This same approach will be used for experimental data; recall that $\epsilon_\mathrm{fit}$ will be available for reconstructions of experimental data but $\epsilon_\mathrm{recons}$ cannot be known. Using a higher number of free parameters will dramatically increase the computational expense and also increase the possibility of overfitting the data, leading to worse estimates.

Further simulations were conducted using the same beam parameters, but reducing the energy from 500~MeV to 42~MeV to more closely align with the experimental scenario at AWA. This change will reduce the validity of the thin lens approximation, as the less rigid particles move during the wakefield interaction, causing both the particle distribution and wake distribution to evolve along the structure. In this case, the virtual screen is 0.2~m downstream of the end of the structure. But, since we now must assume that the beam evolves within the structure the effective drift, $L$, also includes one half of the structure 15~cm structure length, \textit{i.e.} $L$~=~0.28~m. We still assume that the evolution within the structure is sufficiently small that the wakefields do not change. As expected, the lower energy beam does reduce the accuracy of the model (summarized in Table \ref{tab:resultsSummary}). Crucially though, since the difference from the ground truth, $\epsilon_\mathrm{recons}$, is only 4.3\% at 42~MeV, this is still a very high-quality reconstruction and illustrates the flexibility of this approach and its applicability to experimental scenarios of interest. In a later section, we will discuss some possible extensions to the model which could improve its reconstructions further from the thin lens limit.

\begin{figure}[tb] 
\centering
\includegraphics[width=\linewidth]{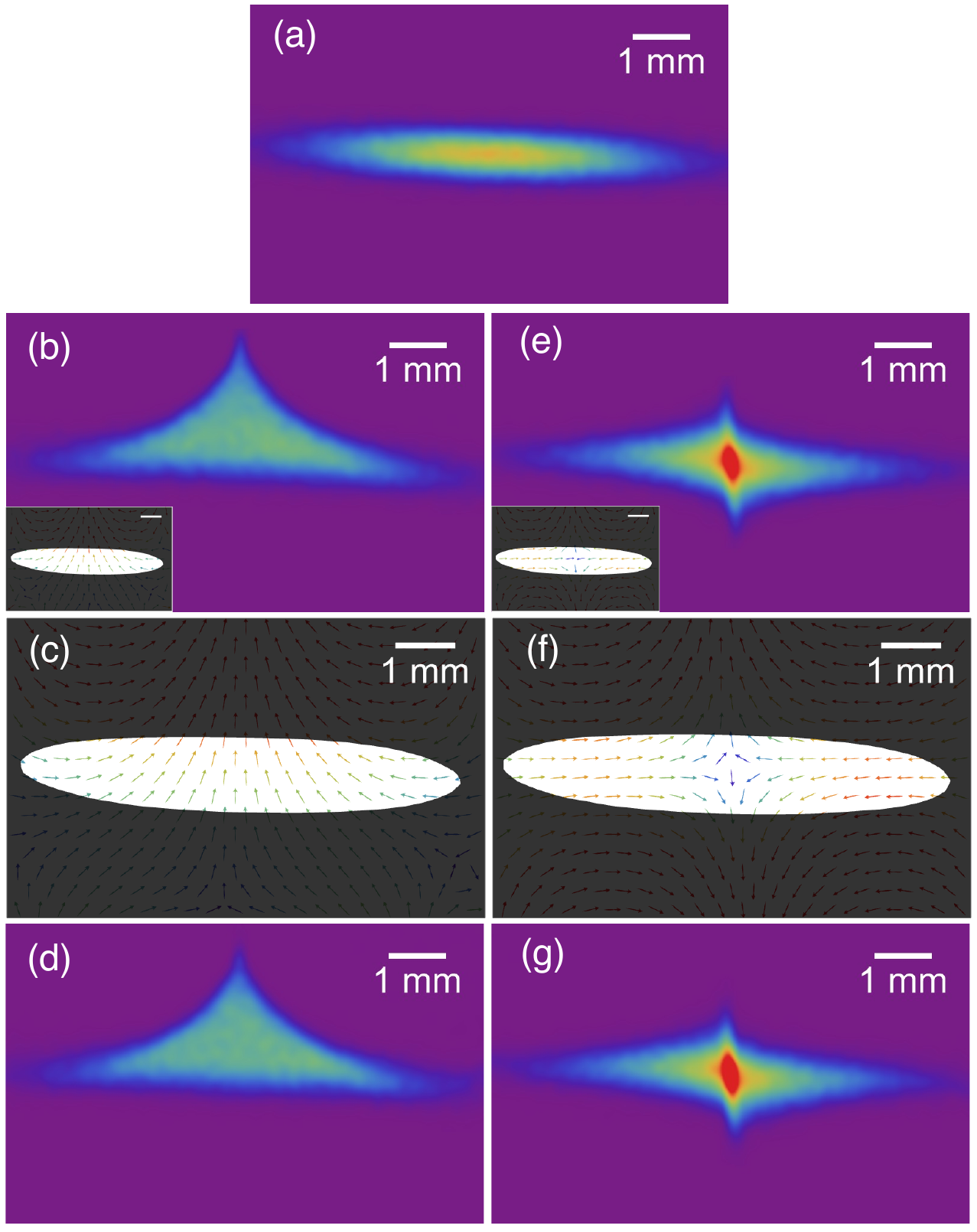}
\caption{Reconstruction of without-structure beam, \textbf{(a)},  into different with-structure beams. This illustrates how the technique is single-shot, once a without-structure image has been collected. 
The first row  is the without-structure beam, the second row is the ground truth, with-structure beam, the third row is the reconstructed wakefields, and the final row is the reconstructed with-structure beams.
Both cases start from the same without-structure beam, \textbf{(a)}. The left column, \textbf{(b,c,d)}, corresponds to the \formatCaseName{42 MeV, top slab, without PDF} simulation while the right column, \textbf{(e,f,g)}, corresponds to the \formatCaseName{42 MeV, double slab, without PDF} simulation. The insets on \textbf{b} and \textbf{e} show the ground truth wakefields from each simulation (scale bars = 1~mm).}
\label{fig:multiReconstruction}
\end{figure}

This same 42~MeV simulated beam is shown reconstructed in several ways in Fig.~\ref{fig:multiReconstruction}. In the first case, it is propagated past a single dielectric slab on the top, while in the second case, it passes between a pair of slabs: one on top and one on the bottom. These reconstructions were accomplished using the same without-structure screen image, illustrating the single-shot nature of the technique: once a without-structure image is acquired, provided the beam is stable over time, all subsequent with-structure screen images can be reconstructed on a single-shot basis.

\section{Reconstructing experimental data}

\begin{figure}[tb] 
\centering
\includegraphics[width=1.0\linewidth]{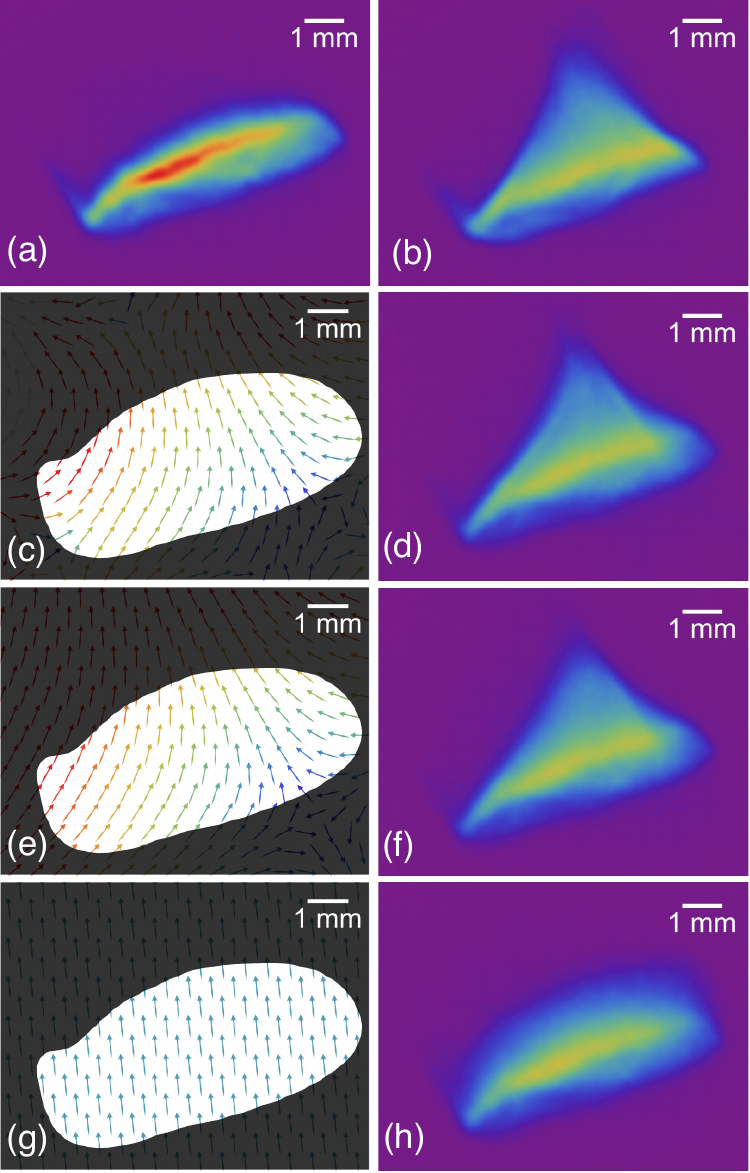}
\caption{Reconstruction of experimental data with \formatCaseName{Top slab, 7.2$^{\circ}$ tilt} beam case under different reconstruction conditions. For additional details refer to Table~\ref{tab:resultsSummary}. \textbf{(a)} Measured without-structure beam. \textbf{(b)} Measured with-structure beam. \textbf{(c-d)} Reconstructed wakefields and implied with-structure beam to $n=6$. \textbf{(e-f)} Reconstructed wakefields and implied with-structure beam to $n=3$. \textbf{(g-h)} Reconstructed wakefields and implied with-structure beam to $n=1$.}
\label{fig:experimentReconstructionConditions}
\end{figure}

This technique has also been applied to experimental data, summarized in Table \ref{tab:resultsSummary}. In Fig.~\ref{fig:experimentReconstructionConditions}, a dielectric wakefield interaction at the AWA between a 7.2$^{\circ}$ tilted, 42 MeV beam has been reconstructed up to $6^\mathrm{th}$ multipole order with lower order reconstructions also shown. This effort is discussed in much greater detail in \cite{lynnSkew} where the reconstructed wakefields are compared to both simulations and analytic models of the skew wake interaction. In this case, the PDF of $\mathcal{K}$ was estimated with Eq.~\ref{eq:1DPDF}. $x_c$ and $y_c$ were also varied, for a total parameter space dimension of 15. The final reconstruction agrees with the actual with-structure beam image to within 6.0\%, suggesting a successful inference of the true wakefields. 

%Additionally, another set of experimental reconstructions was conducted. Here, the wakefields for a single without-structure 42 MeV beam, tilted at 2.2$^{\circ}$, were reconstructed for a variety of with-structure scenarios, shown in Fig.~\ref{fig:multiReconstruction}, with details in Table \ref{tab:resultsSummary}. As with the simulated examples, this illustrates how, once a without-structure image is in hand, it can be used to reconstruct all subsequent with-structure runs.

Naturally, shot-to-shot and long term drifts in the beamline will reduce the quality of the reconstruction. Obviously, if the without-structure reference beam used in the reconstruction algorithm does not remain substantially similar to the actual beam which entered the structure, the reconstruction will be less accurate. The inclusion of free parameters to remove jitter, $x_c$ and $y_c$, is able to address only the most basic sort of jitter, and not longer term drifts.

\section{Discussion}
We have introduced and explained a technique for reconstructing the transverse self-wakefields of a beam without additional experimental infrastructure. The approach has been benchmarked against simulations where the simulated fields may serve as a ground truth, enabling us to quantify the accuracy of the technique. We have, in this way, found agreement at the few-percent level for experimentally relevant scenarios. This technique, with its underlying methods explained in detail, has also been used to reconstruct the transverse self-wakes for experimental data. Comparison to simulation and theory for the particular experiment in question is given elsewhere \cite{lynnSkew}. 

With improvements to the implementation of the technique, specifically in  the numerical optimizer, it may be possible to use this technique as a real-time diagnostic. By combining the reconstructed wakefield information with advanced beam control (\textit{e.g.} multileaf collimators \cite{majernik2023beam} or a range of other options \cite{ha2022bunch}) possibly also leveraging additional non-destructive diagnostic information \cite{oshea2016observation, andonian2012dielectric, roussel2020longitudinal} it should be possible to exert very precise control over the wakes to establish a desired condition.

Although this technique has been demonstrated to be useful for a range of experimentally relevant scenarios, future extensions may further increase its range of applicability. One such step that would further loosen the thin lens requirement would be to reconstruct in multiple steps, evolving the particle distribution (such that each particle now can sample different parts of the wakefield) along the structure but assuming that the wakefield functions themselves are unchanging. It may also be possible to fit multiple $\kappa$ PDFs for different moments to loosen the requirement that the beam be much shorter than all relevant mode wavelengths. 

In a wakefield accelerator, any accelerated bunch will necessarily be far enough from the drive bunch to violate the assumption that the total current profile is short compared to the structure modes. However, it may be possible to reconstruct the wakes for the drive bunch and witness bunch independently. Specifically, this may be approached by running a drive bunch alone with and without the structure present; this technique can be applied directly to reconstruct its self wakes. Then, the drive and witness can be run through the structure and the drive beam's contribution to the screen images may be background subtracted in an effort to capture only the witness beam's with-structure and without-structure distributions. Subsequently, this reconstruction algorithm may be applied to the background-subtracted, witness-only images. This would be exciting future work and could function synergistically with patterned witness beams, \textit{e.g.} \cite{halavanau2021}. By using short witness beams and adjusting the delay relative to the drive beam, it would also be possible to measure the transverse wakefields at different longitudinal positions to reconstruct the full 3D wakefields. %Panofsky-Wenzel

\begin{acknowledgments}
We gratefully acknowledge the valuable support from the staff of the Argonne Wakefield Accelerator and fruitful discussions with S. Baturin. This work was supported by DOE Grants No. DE-SC0017648 and DE-SC0009914. TX and PP are supported by DOE Grant No. DE-SC0022010.
\end{acknowledgments}

\appendix
\section{Summary of reconstructions}

Reconstruction details and figures of merit for selected cases are summarized in Table \ref{tab:resultsSummary}.

\begin{table*}
\setlength\extrarowheight{4pt} %NMM 2023-01-05 - added to pad row height; otherwise looks smushed 
\noindent 

\begin{tabular}{|c|c|c|c|c|c|c|c|}
  \hline
  \textbf{Name} & \textbf{Data type} & $\epsilon_\mathrm{slicewise}$ [\%] & $\epsilon_\mathrm{agg}$ [\%] & \textbf{Multipole} & \textbf{Total free} & $\epsilon_\mathrm{fit}$ [\%] & $\epsilon_\mathrm{recons}$ [\%] \\
  \phantom{xxxxxxxxxxxxxxxxxxxxxxxxxxxxxxxx} & \phantom{xxxxxxxxxxx} &  &  & \textbf{order, $n$} & \textbf{ parameters} &  &  \\ \hline
  \formatCaseName{500 MeV, top slab, with PDF} & \multirow{11}*{Simulation} & \multirow{8}*{1.5} & \multirow{8}*{2.3} & 6 & 12 & 1.3 & 2.7\\ \cline{1-1}\cline{5-8}
  \multirow{7}*{\formatCaseName{500 MeV, top slab, without PDF}} &  &  &  & 6 & 13 & 1.8 & 3.7\\ \cline{5-8}
   &  &  &  & 5 & 11 & 2.1 & 4.0\\ \cline{5-8}
   &  &  &  & 4 & 9 & 4.0 & 9.9\\ \cline{5-8}
   &  &  &  & 3 & 7 & 7.1 & 10\\ \cline{5-8}
   &  &  &  & 2 & 5 & 14 & 24\\ \cline{5-8}
   &  &  &  & 1 & 3 & 20 & 65\\ \cline{5-8}
   &  &  &  & 0 & 0 & 30 & 100\\ \cline{1-1}\cline{3-8}
  \formatCaseName{42 MeV, top slab, with PDF} & &\multirow{2}*{2.4} & \multirow{2}*{3.2} & 6 & 12 & 1.5 & 4.3\\\cline{1-1}\cline{5-8}
  \formatCaseName{42 MeV, top slab, without PDF} & & &  & 6 & 13 & 1.7 & 4.6\\\cline{1-1}\cline{3-8}
  \formatCaseName{42 MeV, double slab, without PDF} & & 2.2 & 2.6 & 6 & 13 & 1.7 & 4.1\\\hline

  \multirow{7}*{\formatCaseName{Top slab, 7.2$^{\circ}$ tilt}} & \multirow{9}*{Experiment}  & \multirow{9}*{N/A} & \multirow{9}*{N/A} & 6 & 15 & 6.0 & \multirow{9}*{N/A} \\ \cline{5-7}
   &  &  &  & 5 & 13 & 6.0 & \\ \cline{5-7}
   &  &  &  & 4 & 11 & 6.2 & \\ \cline{5-7}
   &  &  &  & 3 & 9 & 6.2 & \\ \cline{5-7}
   &  &  &  & 2 & 7 & 6.4 & \\ \cline{5-7}
   &  &  &  & 1 & 5 & 15 & \\ \cline{5-7}
   &  &  &  & 0 & 2 & 25 & \\ \cline{1-1}\cline{5-7}
  \formatCaseName{Bottom slab, 2.2$^{\circ}$ tilt}& & & & 6 & 15 & 8.9 &\\\cline{1-1}\cline{5-7}
   \formatCaseName{Top slab, 2.2$^{\circ}$ tilt} & & & & 6 & 15 & 6.1 &\\\hline

\end{tabular}
\caption{Summary of figures of merit for reconstructions of simulated and experimental screen images. $\epsilon_\mathrm{slicewise}$ (Eq.~\ref{eq:slicewiseError}) compares the slicewise multipole fits to the ground truth wakes. $\epsilon_\mathrm{agg}$ (Eq.~\ref{eq:aggError}) compares the aggregate multipole coefficients and longitudinally dependent multiplicative factor, $\kappa(z)$, to the ground truth wakes. $\epsilon_\mathrm{fit}$ (Eq.~\ref{eq:optimizerError}) is the figure of merit for the optimizer and expresses the error between the binned without-structure beam, transformed according to the best parameter vector, and the with-structure beam. $\epsilon_\mathrm{recons}$ (Eq.~\ref{eq:reconstructionError}) compares the reconstructed wakes to the ground truth wakes.}
\label{tab:resultsSummary}
\end{table*}

\clearpage 

\bibliography{References}% Produces the bibliography via BibTeX.

\begin{thebibliography}{23}%
\makeatletter
\providecommand \@ifxundefined [1]{%
 \@ifx{#1\undefined}
}%
\providecommand \@ifnum [1]{%
 \ifnum #1\expandafter \@firstoftwo
 \else \expandafter \@secondoftwo
 \fi
}%
\providecommand \@ifx [1]{%
 \ifx #1\expandafter \@firstoftwo
 \else \expandafter \@secondoftwo
 \fi
}%
\providecommand \natexlab [1]{#1}%
\providecommand \enquote  [1]{``#1''}%
\providecommand \bibnamefont  [1]{#1}%
\providecommand \bibfnamefont [1]{#1}%
\providecommand \citenamefont [1]{#1}%
\providecommand \href@noop [0]{\@secondoftwo}%
\providecommand \href [0]{\begingroup \@sanitize@url \@href}%
\providecommand \@href[1]{\@@startlink{#1}\@@href}%
\providecommand \@@href[1]{\endgroup#1\@@endlink}%
\providecommand \@sanitize@url [0]{\catcode `\\12\catcode `\$12\catcode
  `\&12\catcode `\#12\catcode `\^12\catcode `\_12\catcode `\%12\relax}%
\providecommand \@@startlink[1]{}%
\providecommand \@@endlink[0]{}%
\providecommand \url  [0]{\begingroup\@sanitize@url \@url }%
\providecommand \@url [1]{\endgroup\@href {#1}{\urlprefix }}%
\providecommand \urlprefix  [0]{URL }%
\providecommand \Eprint [0]{\href }%
\providecommand \doibase [0]{https://doi.org/}%
\providecommand \selectlanguage [0]{\@gobble}%
\providecommand \bibinfo  [0]{\@secondoftwo}%
\providecommand \bibfield  [0]{\@secondoftwo}%
\providecommand \translation [1]{[#1]}%
\providecommand \BibitemOpen [0]{}%
\providecommand \bibitemStop [0]{}%
\providecommand \bibitemNoStop [0]{.\EOS\space}%
\providecommand \EOS [0]{\spacefactor3000\relax}%
\providecommand \BibitemShut  [1]{\csname bibitem#1\endcsname}%
\let\auto@bib@innerbib\@empty
%</preamble>
\bibitem [{\citenamefont {Bettoni}\ \emph {et~al.}(2016)\citenamefont
  {Bettoni}, \citenamefont {Craievich}, \citenamefont {Lutman},\ and\
  \citenamefont {Pedrozzi}}]{bettoni2016}%
  \BibitemOpen
  \bibfield  {author} {\bibinfo {author} {\bibfnamefont {S.}~\bibnamefont
  {Bettoni}}, \bibinfo {author} {\bibfnamefont {P.}~\bibnamefont {Craievich}},
  \bibinfo {author} {\bibfnamefont {A.}~\bibnamefont {Lutman}},\ and\ \bibinfo
  {author} {\bibfnamefont {M.}~\bibnamefont {Pedrozzi}},\ }\bibfield  {title}
  {\bibinfo {title} {Temporal profile measurements of relativistic electron
  bunch based on wakefield generation},\ }\href@noop {} {\bibfield  {journal}
  {\bibinfo  {journal} {Physical Review Accelerators and Beams}\ }\textbf
  {\bibinfo {volume} {19}},\ \bibinfo {pages} {021304} (\bibinfo {year}
  {2016})}\BibitemShut {NoStop}%
\bibitem [{\citenamefont {Lynn}\ \emph
  {et~al.}(2023{\natexlab{a}})\citenamefont {Lynn} \emph {et~al.}}]{lynnIPAC}%
  \BibitemOpen
  \bibfield  {author} {\bibinfo {author} {\bibfnamefont {W.}~\bibnamefont
  {Lynn}} \emph {et~al.},\ }\bibfield  {title} {\bibinfo {title} {Demonstration
  of transverse stability in an alternating symmetry planar dielectric
  structure},\ }in\ \href
  {https://doi.org/https://doi.org/10.18429/JACoW-IPAC-23-TUPA080} {\emph
  {\bibinfo {booktitle} {Proc. IPAC'23}}},\ \bibinfo {series and number}
  {\bibinfo {series} {IPAC'23 - 14th International Particle Accelerator
  Conference}\ No.~\bibinfo {number} {14}}\ (\bibinfo  {publisher} {JACoW
  Publishing, Geneva, Switzerland},\ \bibinfo {year} {2023})\ pp.\ \bibinfo
  {pages} {972--976}\BibitemShut {NoStop}%
\bibitem [{\citenamefont {Lynn}\ \emph {et~al.}(2022)\citenamefont {Lynn},
  \citenamefont {Andonian}, \citenamefont {Doran}, \citenamefont {Kim},
  \citenamefont {Majernik}, \citenamefont {OTool}, \citenamefont {Piot},
  \citenamefont {Power}, \citenamefont {Rosenzweig},\ and\ \citenamefont
  {Wisniewski}}]{lynnNAPAC}%
  \BibitemOpen
  \bibfield  {author} {\bibinfo {author} {\bibfnamefont {W.}~\bibnamefont
  {Lynn}}, \bibinfo {author} {\bibfnamefont {G.}~\bibnamefont {Andonian}},
  \bibinfo {author} {\bibfnamefont {D.}~\bibnamefont {Doran}}, \bibinfo
  {author} {\bibfnamefont {S.}~\bibnamefont {Kim}}, \bibinfo {author}
  {\bibfnamefont {N.}~\bibnamefont {Majernik}}, \bibinfo {author}
  {\bibfnamefont {S.}~\bibnamefont {OTool}}, \bibinfo {author} {\bibfnamefont
  {P.}~\bibnamefont {Piot}}, \bibinfo {author} {\bibfnamefont {J.}~\bibnamefont
  {Power}}, \bibinfo {author} {\bibfnamefont {J.}~\bibnamefont {Rosenzweig}},\
  and\ \bibinfo {author} {\bibfnamefont {E.}~\bibnamefont {Wisniewski}},\
  }\bibfield  {title} {\bibinfo {title} {Transverse stability in an alternating
  symmetry planar dielectric wakefield structure},\ }\bibfield  {journal}
  {\bibinfo  {journal} {JACoW}\ }\href
  {https://doi.org/10.18429/JACoW-NAPAC2022-TUPA82}
  {10.18429/JACoW-NAPAC2022-TUPA82} (\bibinfo {year} {2022})\BibitemShut
  {NoStop}%
\bibitem [{\citenamefont {Li}\ \emph {et~al.}(2014)\citenamefont {Li},
  \citenamefont {Gai}, \citenamefont {Jing}, \citenamefont {Power},
  \citenamefont {Tang},\ and\ \citenamefont {Zholents}}]{li2014}%
  \BibitemOpen
  \bibfield  {author} {\bibinfo {author} {\bibfnamefont {C.}~\bibnamefont
  {Li}}, \bibinfo {author} {\bibfnamefont {W.}~\bibnamefont {Gai}}, \bibinfo
  {author} {\bibfnamefont {C.}~\bibnamefont {Jing}}, \bibinfo {author}
  {\bibfnamefont {J.}~\bibnamefont {Power}}, \bibinfo {author} {\bibfnamefont
  {C.}~\bibnamefont {Tang}},\ and\ \bibinfo {author} {\bibfnamefont
  {A.}~\bibnamefont {Zholents}},\ }\bibfield  {title} {\bibinfo {title} {High
  gradient limits due to single bunch beam breakup in a collinear dielectric
  wakefield accelerator},\ }\href@noop {} {\bibfield  {journal} {\bibinfo
  {journal} {Physical Review Special Topics-Accelerators and Beams}\ }\textbf
  {\bibinfo {volume} {17}},\ \bibinfo {pages} {091302} (\bibinfo {year}
  {2014})}\BibitemShut {NoStop}%
\bibitem [{\citenamefont {O’Shea}\ \emph {et~al.}(2020)\citenamefont
  {O’Shea}, \citenamefont {Andonian}, \citenamefont {Baturin}, \citenamefont
  {Clarke}, \citenamefont {Hoang}, \citenamefont {Hogan}, \citenamefont
  {Naranjo}, \citenamefont {Williams}, \citenamefont {Yakimenko},\ and\
  \citenamefont {Rosenzweig}}]{oshea2020}%
  \BibitemOpen
  \bibfield  {author} {\bibinfo {author} {\bibfnamefont {B.~D.}\ \bibnamefont
  {O’Shea}}, \bibinfo {author} {\bibfnamefont {G.}~\bibnamefont {Andonian}},
  \bibinfo {author} {\bibfnamefont {S.}~\bibnamefont {Baturin}}, \bibinfo
  {author} {\bibfnamefont {C.~I.}\ \bibnamefont {Clarke}}, \bibinfo {author}
  {\bibfnamefont {P.}~\bibnamefont {Hoang}}, \bibinfo {author} {\bibfnamefont
  {M.~J.}\ \bibnamefont {Hogan}}, \bibinfo {author} {\bibfnamefont
  {B.}~\bibnamefont {Naranjo}}, \bibinfo {author} {\bibfnamefont {O.~B.}\
  \bibnamefont {Williams}}, \bibinfo {author} {\bibfnamefont {V.}~\bibnamefont
  {Yakimenko}},\ and\ \bibinfo {author} {\bibfnamefont {J.~B.}\ \bibnamefont
  {Rosenzweig}},\ }\bibfield  {title} {\bibinfo {title} {Suppression of
  deflecting forces in planar-symmetric dielectric wakefield accelerating
  structures with elliptical bunches},\ }\href@noop {} {\bibfield  {journal}
  {\bibinfo  {journal} {Physical review letters}\ }\textbf {\bibinfo {volume}
  {124}},\ \bibinfo {pages} {104801} (\bibinfo {year} {2020})}\BibitemShut
  {NoStop}%
\bibitem [{\citenamefont {Majernik}\ \emph {et~al.}(2022)\citenamefont
  {Majernik}, \citenamefont {Andonian}, \citenamefont {Williams}, \citenamefont
  {O'Shea}, \citenamefont {Hoang}, \citenamefont {Clarke}, \citenamefont
  {Hogan}, \citenamefont {Yakimenko},\ and\ \citenamefont
  {Rosenzweig}}]{majernik2022}%
  \BibitemOpen
  \bibfield  {author} {\bibinfo {author} {\bibfnamefont {N.}~\bibnamefont
  {Majernik}}, \bibinfo {author} {\bibfnamefont {G.}~\bibnamefont {Andonian}},
  \bibinfo {author} {\bibfnamefont {O.}~\bibnamefont {Williams}}, \bibinfo
  {author} {\bibfnamefont {B.}~\bibnamefont {O'Shea}}, \bibinfo {author}
  {\bibfnamefont {P.}~\bibnamefont {Hoang}}, \bibinfo {author} {\bibfnamefont
  {C.}~\bibnamefont {Clarke}}, \bibinfo {author} {\bibfnamefont
  {M.}~\bibnamefont {Hogan}}, \bibinfo {author} {\bibfnamefont
  {V.}~\bibnamefont {Yakimenko}},\ and\ \bibinfo {author} {\bibfnamefont
  {J.}~\bibnamefont {Rosenzweig}},\ }\bibfield  {title} {\bibinfo {title}
  {Positron driven high-field terahertz waves via dielectric wakefield
  interaction},\ }\href@noop {} {\bibfield  {journal} {\bibinfo  {journal}
  {Physical Review Research}\ }\textbf {\bibinfo {volume} {4}},\ \bibinfo
  {pages} {023065} (\bibinfo {year} {2022})}\BibitemShut {NoStop}%
\bibitem [{\citenamefont {Marques}\ \emph {et~al.}(1996)\citenamefont
  {Marques}, \citenamefont {Geindre}, \citenamefont {Amiranoff}, \citenamefont
  {Audebert}, \citenamefont {Gauthier}, \citenamefont {Antonetti},\ and\
  \citenamefont {Grillon}}]{marques1996temporal}%
  \BibitemOpen
  \bibfield  {author} {\bibinfo {author} {\bibfnamefont {J.}~\bibnamefont
  {Marques}}, \bibinfo {author} {\bibfnamefont {J.}~\bibnamefont {Geindre}},
  \bibinfo {author} {\bibfnamefont {F.}~\bibnamefont {Amiranoff}}, \bibinfo
  {author} {\bibfnamefont {P.}~\bibnamefont {Audebert}}, \bibinfo {author}
  {\bibfnamefont {J.}~\bibnamefont {Gauthier}}, \bibinfo {author}
  {\bibfnamefont {A.}~\bibnamefont {Antonetti}},\ and\ \bibinfo {author}
  {\bibfnamefont {G.}~\bibnamefont {Grillon}},\ }\bibfield  {title} {\bibinfo
  {title} {Temporal and spatial measurements of the electron density
  perturbation produced in the wake of an ultrashort laser pulse},\ }\href@noop
  {} {\bibfield  {journal} {\bibinfo  {journal} {Physical review letters}\
  }\textbf {\bibinfo {volume} {76}},\ \bibinfo {pages} {3566} (\bibinfo {year}
  {1996})}\BibitemShut {NoStop}%
\bibitem [{\citenamefont {Siders}\ \emph {et~al.}(1996)\citenamefont {Siders},
  \citenamefont {Le~Blanc}, \citenamefont {Fisher}, \citenamefont {Tajima},
  \citenamefont {Downer}, \citenamefont {Babine}, \citenamefont {Stepanov},\
  and\ \citenamefont {Sergeev}}]{siders1996laser}%
  \BibitemOpen
  \bibfield  {author} {\bibinfo {author} {\bibfnamefont {C.}~\bibnamefont
  {Siders}}, \bibinfo {author} {\bibfnamefont {S.}~\bibnamefont {Le~Blanc}},
  \bibinfo {author} {\bibfnamefont {D.}~\bibnamefont {Fisher}}, \bibinfo
  {author} {\bibfnamefont {T.}~\bibnamefont {Tajima}}, \bibinfo {author}
  {\bibfnamefont {M.}~\bibnamefont {Downer}}, \bibinfo {author} {\bibfnamefont
  {A.}~\bibnamefont {Babine}}, \bibinfo {author} {\bibfnamefont
  {A.}~\bibnamefont {Stepanov}},\ and\ \bibinfo {author} {\bibfnamefont
  {A.}~\bibnamefont {Sergeev}},\ }\bibfield  {title} {\bibinfo {title} {Laser
  wakefield excitation and measurement by femtosecond longitudinal
  interferometry},\ }\href@noop {} {\bibfield  {journal} {\bibinfo  {journal}
  {Physical review letters}\ }\textbf {\bibinfo {volume} {76}},\ \bibinfo
  {pages} {3570} (\bibinfo {year} {1996})}\BibitemShut {NoStop}%
\bibitem [{\citenamefont {Matlis}\ \emph {et~al.}(2006)\citenamefont {Matlis},
  \citenamefont {Reed}, \citenamefont {Bulanov}, \citenamefont {Chvykov},
  \citenamefont {Kalintchenko}, \citenamefont {Matsuoka}, \citenamefont
  {Rousseau}, \citenamefont {Yanovsky}, \citenamefont {Maksimchuk},
  \citenamefont {Kalmykov} \emph {et~al.}}]{matlis2006snapshots}%
  \BibitemOpen
  \bibfield  {author} {\bibinfo {author} {\bibfnamefont {N.~H.}\ \bibnamefont
  {Matlis}}, \bibinfo {author} {\bibfnamefont {S.}~\bibnamefont {Reed}},
  \bibinfo {author} {\bibfnamefont {S.~S.}\ \bibnamefont {Bulanov}}, \bibinfo
  {author} {\bibfnamefont {V.}~\bibnamefont {Chvykov}}, \bibinfo {author}
  {\bibfnamefont {G.}~\bibnamefont {Kalintchenko}}, \bibinfo {author}
  {\bibfnamefont {T.}~\bibnamefont {Matsuoka}}, \bibinfo {author}
  {\bibfnamefont {P.}~\bibnamefont {Rousseau}}, \bibinfo {author}
  {\bibfnamefont {V.}~\bibnamefont {Yanovsky}}, \bibinfo {author}
  {\bibfnamefont {A.}~\bibnamefont {Maksimchuk}}, \bibinfo {author}
  {\bibfnamefont {S.}~\bibnamefont {Kalmykov}}, \emph {et~al.},\ }\bibfield
  {title} {\bibinfo {title} {Snapshots of laser wakefields},\ }\href@noop {}
  {\bibfield  {journal} {\bibinfo  {journal} {Nature Physics}\ }\textbf
  {\bibinfo {volume} {2}},\ \bibinfo {pages} {749} (\bibinfo {year}
  {2006})}\BibitemShut {NoStop}%
\bibitem [{\citenamefont {Buck}\ \emph {et~al.}(2011)\citenamefont {Buck},
  \citenamefont {Nicolai}, \citenamefont {Schmid}, \citenamefont {Sears},
  \citenamefont {S{\"a}vert}, \citenamefont {Mikhailova}, \citenamefont
  {Krausz}, \citenamefont {Kaluza},\ and\ \citenamefont
  {Veisz}}]{buck2011real}%
  \BibitemOpen
  \bibfield  {author} {\bibinfo {author} {\bibfnamefont {A.}~\bibnamefont
  {Buck}}, \bibinfo {author} {\bibfnamefont {M.}~\bibnamefont {Nicolai}},
  \bibinfo {author} {\bibfnamefont {K.}~\bibnamefont {Schmid}}, \bibinfo
  {author} {\bibfnamefont {C.}~\bibnamefont {Sears}}, \bibinfo {author}
  {\bibfnamefont {A.}~\bibnamefont {S{\"a}vert}}, \bibinfo {author}
  {\bibfnamefont {J.~M.}\ \bibnamefont {Mikhailova}}, \bibinfo {author}
  {\bibfnamefont {F.}~\bibnamefont {Krausz}}, \bibinfo {author} {\bibfnamefont
  {M.~C.}\ \bibnamefont {Kaluza}},\ and\ \bibinfo {author} {\bibfnamefont
  {L.}~\bibnamefont {Veisz}},\ }\bibfield  {title} {\bibinfo {title} {Real-time
  observation of laser-driven electron acceleration},\ }\href@noop {}
  {\bibfield  {journal} {\bibinfo  {journal} {Nature Physics}\ }\textbf
  {\bibinfo {volume} {7}},\ \bibinfo {pages} {543} (\bibinfo {year}
  {2011})}\BibitemShut {NoStop}%
\bibitem [{\citenamefont {S{\"a}vert}\ \emph {et~al.}(2015)\citenamefont
  {S{\"a}vert}, \citenamefont {Mangles}, \citenamefont {Schnell}, \citenamefont
  {Siminos}, \citenamefont {Cole}, \citenamefont {Leier}, \citenamefont
  {Reuter}, \citenamefont {Schwab}, \citenamefont {M{\"o}ller}, \citenamefont
  {Poder} \emph {et~al.}}]{savert2015direct}%
  \BibitemOpen
  \bibfield  {author} {\bibinfo {author} {\bibfnamefont {A.}~\bibnamefont
  {S{\"a}vert}}, \bibinfo {author} {\bibfnamefont {S.}~\bibnamefont {Mangles}},
  \bibinfo {author} {\bibfnamefont {M.}~\bibnamefont {Schnell}}, \bibinfo
  {author} {\bibfnamefont {E.}~\bibnamefont {Siminos}}, \bibinfo {author}
  {\bibfnamefont {J.}~\bibnamefont {Cole}}, \bibinfo {author} {\bibfnamefont
  {M.}~\bibnamefont {Leier}}, \bibinfo {author} {\bibfnamefont
  {M.}~\bibnamefont {Reuter}}, \bibinfo {author} {\bibfnamefont
  {M.}~\bibnamefont {Schwab}}, \bibinfo {author} {\bibfnamefont
  {M.}~\bibnamefont {M{\"o}ller}}, \bibinfo {author} {\bibfnamefont
  {K.}~\bibnamefont {Poder}}, \emph {et~al.},\ }\bibfield  {title} {\bibinfo
  {title} {Direct observation of the injection dynamics of a laser wakefield
  accelerator using few-femtosecond shadowgraphy},\ }\href@noop {} {\bibfield
  {journal} {\bibinfo  {journal} {Physical review letters}\ }\textbf {\bibinfo
  {volume} {115}},\ \bibinfo {pages} {055002} (\bibinfo {year}
  {2015})}\BibitemShut {NoStop}%
\bibitem [{\citenamefont {Zhang}\ \emph {et~al.}(2016)\citenamefont {Zhang},
  \citenamefont {Hua}, \citenamefont {Xu}, \citenamefont {Li}, \citenamefont
  {Pai}, \citenamefont {Wan}, \citenamefont {Wu}, \citenamefont {Gu},
  \citenamefont {Mori}, \citenamefont {Joshi} \emph
  {et~al.}}]{zhang2016capturing}%
  \BibitemOpen
  \bibfield  {author} {\bibinfo {author} {\bibfnamefont {C.}~\bibnamefont
  {Zhang}}, \bibinfo {author} {\bibfnamefont {J.}~\bibnamefont {Hua}}, \bibinfo
  {author} {\bibfnamefont {X.}~\bibnamefont {Xu}}, \bibinfo {author}
  {\bibfnamefont {F.}~\bibnamefont {Li}}, \bibinfo {author} {\bibfnamefont
  {C.-H.}\ \bibnamefont {Pai}}, \bibinfo {author} {\bibfnamefont
  {Y.}~\bibnamefont {Wan}}, \bibinfo {author} {\bibfnamefont {Y.}~\bibnamefont
  {Wu}}, \bibinfo {author} {\bibfnamefont {Y.}~\bibnamefont {Gu}}, \bibinfo
  {author} {\bibfnamefont {W.}~\bibnamefont {Mori}}, \bibinfo {author}
  {\bibfnamefont {C.}~\bibnamefont {Joshi}}, \emph {et~al.},\ }\bibfield
  {title} {\bibinfo {title} {Capturing relativistic wakefield structures in
  plasmas using ultrashort high-energy electrons as a probe},\ }\href@noop {}
  {\bibfield  {journal} {\bibinfo  {journal} {Scientific reports}\ }\textbf
  {\bibinfo {volume} {6}},\ \bibinfo {pages} {1} (\bibinfo {year}
  {2016})}\BibitemShut {NoStop}%
\bibitem [{\citenamefont {Halavanau}\ \emph {et~al.}(2021)\citenamefont
  {Halavanau}, \citenamefont {Mayes}, \citenamefont {Gessner},\ and\
  \citenamefont {Rosenzweig}}]{halavanau2021}%
  \BibitemOpen
  \bibfield  {author} {\bibinfo {author} {\bibfnamefont {A.}~\bibnamefont
  {Halavanau}}, \bibinfo {author} {\bibfnamefont {C.}~\bibnamefont {Mayes}},
  \bibinfo {author} {\bibfnamefont {S.}~\bibnamefont {Gessner}},\ and\ \bibinfo
  {author} {\bibfnamefont {J.}~\bibnamefont {Rosenzweig}},\ }\bibfield  {title}
  {\bibinfo {title} {Hollow and flat electron beam generation at {FACET-II}},\
  }in\ \href@noop {} {\emph {\bibinfo {booktitle} {12th Int. Particle
  Accelerator Conf.(IPAC’21), Campinas, Brazil}}}\ (\bibinfo {year}
  {2021})\BibitemShut {NoStop}%
\bibitem [{\citenamefont {Tanabe}(2005)}]{tanabe2005}%
  \BibitemOpen
  \bibfield  {author} {\bibinfo {author} {\bibfnamefont {J.~T.}\ \bibnamefont
  {Tanabe}},\ }\href@noop {} {\emph {\bibinfo {title} {Iron dominated
  electromagnets: design, fabrication, assembly and measurements}}}\ (\bibinfo
  {publisher} {World Scientific Publishing Company},\ \bibinfo {year}
  {2005})\BibitemShut {NoStop}%
\bibitem [{\citenamefont {Koziel}\ and\ \citenamefont
  {Yang}(2011)}]{koziel2011}%
  \BibitemOpen
  \bibfield  {author} {\bibinfo {author} {\bibfnamefont {S.}~\bibnamefont
  {Koziel}}\ and\ \bibinfo {author} {\bibfnamefont {X.-S.}\ \bibnamefont
  {Yang}},\ }\href@noop {} {\emph {\bibinfo {title} {Computational
  optimization, methods and algorithms}}},\ Vol.\ \bibinfo {volume} {356}\
  (\bibinfo  {publisher} {Springer},\ \bibinfo {year} {2011})\BibitemShut
  {NoStop}%
\bibitem [{\citenamefont {Roussel}\ \emph {et~al.}(2022)\citenamefont
  {Roussel}, \citenamefont {Edelen}, \citenamefont {Mayes}, \citenamefont
  {Ratner}, \citenamefont {Gonzalez-Aguilera}, \citenamefont {Kim},
  \citenamefont {Wisniewski},\ and\ \citenamefont {Power}}]{roussel2022phase}%
  \BibitemOpen
  \bibfield  {author} {\bibinfo {author} {\bibfnamefont {R.}~\bibnamefont
  {Roussel}}, \bibinfo {author} {\bibfnamefont {A.}~\bibnamefont {Edelen}},
  \bibinfo {author} {\bibfnamefont {C.}~\bibnamefont {Mayes}}, \bibinfo
  {author} {\bibfnamefont {D.}~\bibnamefont {Ratner}}, \bibinfo {author}
  {\bibfnamefont {J.~P.}\ \bibnamefont {Gonzalez-Aguilera}}, \bibinfo {author}
  {\bibfnamefont {S.}~\bibnamefont {Kim}}, \bibinfo {author} {\bibfnamefont
  {E.}~\bibnamefont {Wisniewski}},\ and\ \bibinfo {author} {\bibfnamefont
  {J.}~\bibnamefont {Power}},\ }\bibfield  {title} {\bibinfo {title} {Phase
  space reconstruction from accelerator beam measurements using neural networks
  and differentiable simulations},\ }\href@noop {} {\bibfield  {journal}
  {\bibinfo  {journal} {arXiv preprint arXiv:2209.04505}\ } (\bibinfo {year}
  {2022})}\BibitemShut {NoStop}%
\bibitem [{\citenamefont {Vay}\ \emph {et~al.}(2012)\citenamefont {Vay},
  \citenamefont {Grote}, \citenamefont {Cohen},\ and\ \citenamefont
  {Friedman}}]{WARP}%
  \BibitemOpen
  \bibfield  {author} {\bibinfo {author} {\bibfnamefont {J.-L.}\ \bibnamefont
  {Vay}}, \bibinfo {author} {\bibfnamefont {D.~P.}\ \bibnamefont {Grote}},
  \bibinfo {author} {\bibfnamefont {R.~H.}\ \bibnamefont {Cohen}},\ and\
  \bibinfo {author} {\bibfnamefont {A.}~\bibnamefont {Friedman}},\ }\bibfield
  {title} {\bibinfo {title} {Novel methods in the particle-in-cell accelerator
  code-framework warp},\ }\href {https://doi.org/10.1088/1749-4699/5/1/014019}
  {\bibfield  {journal} {\bibinfo  {journal} {Computational Science \&
  Discovery}\ }\textbf {\bibinfo {volume} {5}},\ \bibinfo {pages} {014019}
  (\bibinfo {year} {2012})}\BibitemShut {NoStop}%
\bibitem [{\citenamefont {Lynn}\ \emph
  {et~al.}(2023{\natexlab{b}})\citenamefont {Lynn} \emph {et~al.}}]{lynnSkew}%
  \BibitemOpen
  \bibfield  {author} {\bibinfo {author} {\bibfnamefont {W.}~\bibnamefont
  {Lynn}} \emph {et~al.},\ }\bibfield  {title} {\bibinfo {title} {Observation
  of skewed electromagnetic wakefields in an asymmetric structure driven by
  flat electron bunches},\ }\href@noop {} {\bibfield  {journal} {\bibinfo
  {journal} {\textit{Submitted}}\ } (\bibinfo {year}
  {2023}{\natexlab{b}})}\BibitemShut {NoStop}%
\bibitem [{\citenamefont {Majernik}\ \emph {et~al.}(2023)\citenamefont
  {Majernik}, \citenamefont {Andonian}, \citenamefont {Lynn}, \citenamefont
  {Kim}, \citenamefont {Lorch}, \citenamefont {Roussel}, \citenamefont {Doran},
  \citenamefont {Wisniewski}, \citenamefont {Whiteford}, \citenamefont {Piot}
  \emph {et~al.}}]{majernik2023beam}%
  \BibitemOpen
  \bibfield  {author} {\bibinfo {author} {\bibfnamefont {N.}~\bibnamefont
  {Majernik}}, \bibinfo {author} {\bibfnamefont {G.}~\bibnamefont {Andonian}},
  \bibinfo {author} {\bibfnamefont {W.}~\bibnamefont {Lynn}}, \bibinfo {author}
  {\bibfnamefont {S.}~\bibnamefont {Kim}}, \bibinfo {author} {\bibfnamefont
  {C.}~\bibnamefont {Lorch}}, \bibinfo {author} {\bibfnamefont
  {R.}~\bibnamefont {Roussel}}, \bibinfo {author} {\bibfnamefont
  {S.}~\bibnamefont {Doran}}, \bibinfo {author} {\bibfnamefont
  {E.}~\bibnamefont {Wisniewski}}, \bibinfo {author} {\bibfnamefont
  {C.}~\bibnamefont {Whiteford}}, \bibinfo {author} {\bibfnamefont
  {P.}~\bibnamefont {Piot}}, \emph {et~al.},\ }\bibfield  {title} {\bibinfo
  {title} {Beam shaping using an ultrahigh vacuum multileaf collimator and
  emittance exchange beamline},\ }\href@noop {} {\bibfield  {journal} {\bibinfo
   {journal} {Physical Review Accelerators and Beams}\ }\textbf {\bibinfo
  {volume} {26}},\ \bibinfo {pages} {022801} (\bibinfo {year}
  {2023})}\BibitemShut {NoStop}%
\bibitem [{\citenamefont {Ha}\ \emph {et~al.}(2022)\citenamefont {Ha},
  \citenamefont {Kim}, \citenamefont {Power}, \citenamefont {Sun},
  \citenamefont {Piot} \emph {et~al.}}]{ha2022bunch}%
  \BibitemOpen
  \bibfield  {author} {\bibinfo {author} {\bibfnamefont {G.}~\bibnamefont
  {Ha}}, \bibinfo {author} {\bibfnamefont {K.-J.}\ \bibnamefont {Kim}},
  \bibinfo {author} {\bibfnamefont {J.}~\bibnamefont {Power}}, \bibinfo
  {author} {\bibfnamefont {Y.}~\bibnamefont {Sun}}, \bibinfo {author}
  {\bibfnamefont {P.}~\bibnamefont {Piot}}, \emph {et~al.},\ }\bibfield
  {title} {\bibinfo {title} {Bunch shaping in electron linear accelerators},\
  }\href@noop {} {\bibfield  {journal} {\bibinfo  {journal} {Reviews of Modern
  Physics}\ }\textbf {\bibinfo {volume} {94}},\ \bibinfo {pages} {025006}
  (\bibinfo {year} {2022})}\BibitemShut {NoStop}%
\bibitem [{\citenamefont {O’Shea}\ \emph {et~al.}(2016)\citenamefont
  {O’Shea}, \citenamefont {Andonian}, \citenamefont {Barber}, \citenamefont
  {Fitzmorris}, \citenamefont {Hakimi}, \citenamefont {Harrison}, \citenamefont
  {Hoang}, \citenamefont {Hogan}, \citenamefont {Naranjo}, \citenamefont
  {Williams} \emph {et~al.}}]{oshea2016observation}%
  \BibitemOpen
  \bibfield  {author} {\bibinfo {author} {\bibfnamefont {B.}~\bibnamefont
  {O’Shea}}, \bibinfo {author} {\bibfnamefont {G.}~\bibnamefont {Andonian}},
  \bibinfo {author} {\bibfnamefont {S.}~\bibnamefont {Barber}}, \bibinfo
  {author} {\bibfnamefont {K.}~\bibnamefont {Fitzmorris}}, \bibinfo {author}
  {\bibfnamefont {S.}~\bibnamefont {Hakimi}}, \bibinfo {author} {\bibfnamefont
  {J.}~\bibnamefont {Harrison}}, \bibinfo {author} {\bibfnamefont
  {P.}~\bibnamefont {Hoang}}, \bibinfo {author} {\bibfnamefont
  {M.}~\bibnamefont {Hogan}}, \bibinfo {author} {\bibfnamefont
  {B.}~\bibnamefont {Naranjo}}, \bibinfo {author} {\bibfnamefont
  {O.}~\bibnamefont {Williams}}, \emph {et~al.},\ }\bibfield  {title} {\bibinfo
  {title} {Observation of acceleration and deceleration in
  gigaelectron-volt-per-metre gradient dielectric wakefield accelerators},\
  }\href@noop {} {\bibfield  {journal} {\bibinfo  {journal} {Nature
  communications}\ }\textbf {\bibinfo {volume} {7}},\ \bibinfo {pages} {12763}
  (\bibinfo {year} {2016})}\BibitemShut {NoStop}%
\bibitem [{\citenamefont {Andonian}\ \emph {et~al.}(2012)\citenamefont
  {Andonian}, \citenamefont {Stratakis}, \citenamefont {Babzien}, \citenamefont
  {Barber}, \citenamefont {Fedurin}, \citenamefont {Hemsing}, \citenamefont
  {Kusche}, \citenamefont {Muggli}, \citenamefont {O’Shea}, \citenamefont
  {Wei} \emph {et~al.}}]{andonian2012dielectric}%
  \BibitemOpen
  \bibfield  {author} {\bibinfo {author} {\bibfnamefont {G.}~\bibnamefont
  {Andonian}}, \bibinfo {author} {\bibfnamefont {D.}~\bibnamefont {Stratakis}},
  \bibinfo {author} {\bibfnamefont {M.}~\bibnamefont {Babzien}}, \bibinfo
  {author} {\bibfnamefont {S.}~\bibnamefont {Barber}}, \bibinfo {author}
  {\bibfnamefont {M.}~\bibnamefont {Fedurin}}, \bibinfo {author} {\bibfnamefont
  {E.}~\bibnamefont {Hemsing}}, \bibinfo {author} {\bibfnamefont
  {K.}~\bibnamefont {Kusche}}, \bibinfo {author} {\bibfnamefont
  {P.}~\bibnamefont {Muggli}}, \bibinfo {author} {\bibfnamefont
  {B.}~\bibnamefont {O’Shea}}, \bibinfo {author} {\bibfnamefont
  {X.}~\bibnamefont {Wei}}, \emph {et~al.},\ }\bibfield  {title} {\bibinfo
  {title} {Dielectric wakefield acceleration of a relativistic electron beam in
  a slab-symmetric dielectric lined waveguide},\ }\href@noop {} {\bibfield
  {journal} {\bibinfo  {journal} {Physical review letters}\ }\textbf {\bibinfo
  {volume} {108}},\ \bibinfo {pages} {244801} (\bibinfo {year}
  {2012})}\BibitemShut {NoStop}%
\bibitem [{\citenamefont {Roussel}\ \emph {et~al.}(2020)\citenamefont
  {Roussel}, \citenamefont {Andonian}, \citenamefont {Rosenzweig},\ and\
  \citenamefont {Baturin}}]{roussel2020longitudinal}%
  \BibitemOpen
  \bibfield  {author} {\bibinfo {author} {\bibfnamefont {R.}~\bibnamefont
  {Roussel}}, \bibinfo {author} {\bibfnamefont {G.}~\bibnamefont {Andonian}},
  \bibinfo {author} {\bibfnamefont {J.}~\bibnamefont {Rosenzweig}},\ and\
  \bibinfo {author} {\bibfnamefont {S.}~\bibnamefont {Baturin}},\ }\bibfield
  {title} {\bibinfo {title} {Longitudinal current profile reconstruction from a
  wakefield response in plasmas and structures},\ }\href@noop {} {\bibfield
  {journal} {\bibinfo  {journal} {Physical Review Accelerators and Beams}\
  }\textbf {\bibinfo {volume} {23}},\ \bibinfo {pages} {121303} (\bibinfo
  {year} {2020})}\BibitemShut {NoStop}%
\end{thebibliography}%

\end{document}